\documentclass[preprint,12pt]{elsarticle}

\usepackage{graphicx}
\usepackage{caption}
\usepackage{subcaption}
\usepackage{amssymb}
\usepackage{csquotes}
\usepackage{lineno}
\usepackage[colorlinks=true, allcolors=blue]{hyperref}
\usepackage[dvipsnames]{xcolor}

\journal{Nuclear Physics A}

\begin{document}

\begin{frontmatter}



\title{Developing the Future of Gamma-ray Astrophysics with Monolithic Silicon Pixels}


\author[a,b,g]{Isabella Brewer}
\author[b,c,g]{Michela Negro}
\author[d]{Nicolas Striebig}
\author[b]{Carolyn Kierans}
\author[b]{Regina Caputo}
\author[d]{Richard Leys}
\author[d]{Ivan Peric}
\author[b,f,g]{Henrike Fleischhack}
\author[e]{Jessica Metcalfe}
\author[b]{Jeremy Perkins}

\affiliation[a]{organization={University of Maryland College Park},
            city={College Park},
            state={MD},
            country={USA}}
            
\affiliation[b]{organization={NASA Goddard Space Flight Center},
            city={Greenbelt},
            state={MD},
            country={USA}}

\affiliation[c]{organization={University of Maryland Baltimore County},
            city={Baltimore},
            state={MD},
            country={USA}}
            
\affiliation[d]{organization={ASIC and Detector Laboratory, Karlsruhe Institute of Technology},
            state={Karlsruhe},
            country={Germany}}
            
\affiliation[e]{organization={Argonne National Laboratory},
            city={Lemont},
            state={IL},
            country={USA}}

\affiliation[f]{organization={Department of Physics, Catholic University of America},
            city={Washington, DC, 20064},
            country={USA}}

\affiliation[g]{organization={Center for Research and Exploration in Space Science and Technology, NASA/GSFC},
            city={Greenbelt},
            state={MD},
            country={USA}}           

\begin{abstract}
This paper explores the potential of AstroPix, a project to develop Complementary Metal Oxide Semiconductor (CMOS) pixels for the next generation of space-based high-energy astrophysics experiments. Multimessenger astrophysics is a rapidly developing field whose upcoming missions need support from new detector technology such as AstroPix. ATLASPix, a monolithic silicon detector optimized for the ATLAS particle detector at CERN, is the beginning of the larger AstroPix project. Energy resolution is a driving parameter in the reconstruction of gamma-ray events, and therefore the characterization of ATLASPix energy resolution is the focus of this paper. The intrinsic energy resolution of the detector exceeded our baseline requirements of $<$10\% at \textcolor{black}{60} keV. The digital output of ATLASPix results in energy resolutions insufficient to advance gamma-ray astronomy. However, the results from the intrinsic energy resolution indicate the digital capability of the detector can be redesigned, and the next generation of pixels for the larger AstroPix project have already been constructed. Iterations of AstroPix-type pixels are an exciting new technology candidate to support new space-based missions.
\end{abstract}



\end{frontmatter}


\section{\large Introduction}
\label{introduction}
\noindent

Multimessenger astronomy, a burgeoning field of collaborative astrophysics that will play a leading role in the next decade, encompasses all four types of messengers from extreme events in our universe: those from strong and weak nuclear forces, gravitational forces, and electromagnetic forces \cite{multimessenger_review}. Observing the universe through photons has been the standard technique for many years, and it is only in the last few decades that we have been able to extend our detections to non-photon messengers \cite{multimessenger_review}. Messengers such as cosmic rays, gravitational waves, and neutrinos all contain unique information about the universe; coordinating the detections of differing signals and identifying their corresponding electromagnetic counterparts maximizes our understanding about the astrophysical processes behind them.

All known mutlimessenger events peak in the gamma-ray energy regime, making this area of observation vital to multimessenger astronomy. The potential of multimessenger astronomy in the soft- and medium-energy gamma-ray range, an under-explored region of observation, is the scientific focus of the proposed the explorer class All-sky Medium Energy Gamma-ray Observatory eXplorer (AMEGO-X)\footnote{https://asd.gsfc.nasa.gov/amego-x/} mission and the All-sky Medium Energy Gamma-ray Observatory (AMEGO)\footnote{https://asd.gsfc.nasa.gov/amego/} mission, a probe mission concept that has been submitted to the Astro2020 Decadal Survey \cite{mcenery}. Recent detections of coincident messengers has led to revolutionary discoveries such as the detection of the short gamma-ray burst (sGRB) GRB170817A and gravitational wave event GW170817 \cite{goldstein}, as well as the possible connection of flaring blazar TXS 0506+056 with neutrino IceCube-170922A \cite{coincident_neutrino}. With the expected increased capacity to detect neutrinos (IceCube Gen2 \cite{IceCube_Collab}) and gravitational waves (Advanced LIGO/VIRGO) in the next decade, AMEGO-X and AMEGO recognize the importance of increasing our ability to detect gamma-ray counterparts \cite{Kierans_2020}. 

\subsection{Design considerations for multimessenger gamma-ray telescopes}
\label{sec:design}

\begin{figure}[htb]
\centering
\includegraphics[width=0.6\textwidth]{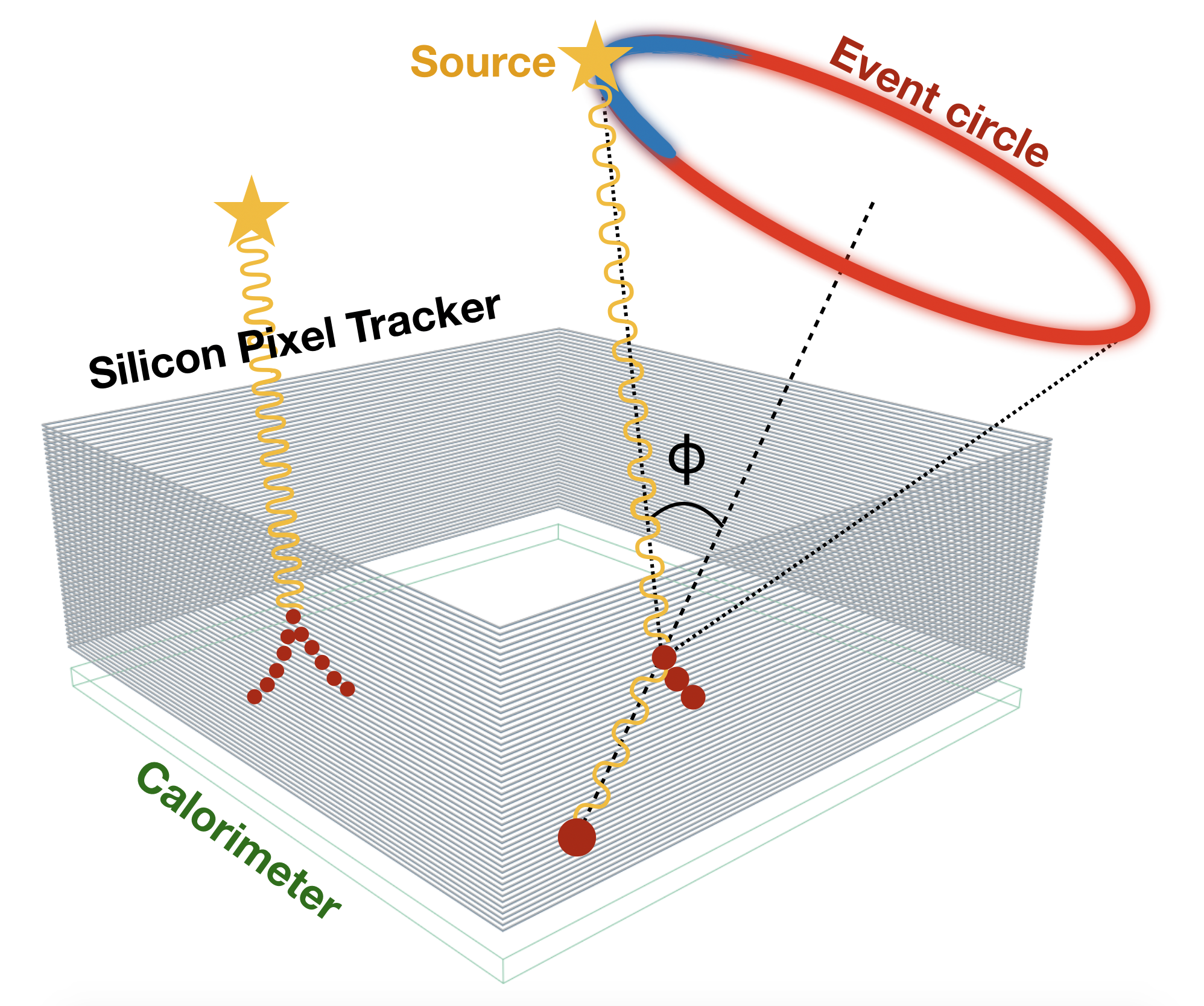}
\caption{\textcolor{black}{The two dominate gamma-ray interactions are shown for a silicon pixel tracker telescope. At energies greater than $\sim$10~MeV, a photon predominately converts to an electron-positron pair, and the initial photon direction can be determined by the tracks within the instrument. In the Compton regime below $\sim$10~MeV, the initial direction of the gamma-ray can be constrained to a circle on the sky with radius defined by the Compton scatter angle $\phi$. If the direction of the Compton-scattered electron can be tracked, then the event circle can be reduced to the blue arc shown.}}
\label{fig:eventschematic}
\end{figure}

In order to enable multimessenger astronomy, missions like AMEGO-X and AMEGO must be sensitive to soft- to medium-energy gamma rays from about 100 keV, in order to detect sGRBs, an electromagnetic counterpart to neutron star mergers, to the hundreds of MeV, to detect counterparts to active galactic nuclei (AGN). To cover this broad range of energies, a multimessenger telescope needs to be sensitive to gamma rays that undergo Compton scattering at lower energies and pair production at higher energies. Detecting two types of photon interactions within a single detector medium places certain design requirements on an instrument. Particle tracking and position resolution are especially important when reconstructing Compton events, which involves tracking the inelastic collision of a photon and an electron. The scattered electron travels a short distance before being photoabsorbed, requiring that the position of the Compton scatter itself has to be measured in three dimensions and within a single layer of a tracking detector. The Compton angle \textcolor{black}{$\phi$} of this inelastic collision is calculated from the recorded energies of the scattered photon and scattered electron. The Compton angle is then used to form a cone whose opening constrains the origin of the incident photon to a circle in the sky\textcolor{black}{, as shown in Figure~\ref{fig:eventschematic}}. To calculate the axis of the Compton cone, a Compton telescope must be capable of providing 3D tracking data with fine position resolution. Reconstruction of Compton events therefore requires precise position and energy resolution of the initial photon, scattered photon, and electron, as large uncertainties in any one measurement increases the uncertainty of the final calculation of the origin of the photon. 

For high energy photons above $\sim$10 MeV, the dominate interaction mechanism is pair production. With pair production, a high-energy photon will convert into an electron and positron pair. The positions of the electron and positron pair have to be traced as they move through the detector, as the particle tracks are used to reconstruct the origin of the incident photon. A common way to achieve the necessary position resolution is with a tracker-type detector comprised of many thin layers of segmented detector material. A calorimeter beneath the tracker contains the electromagnetic shower from the electron-positron pair and provides the total energy of the event. A typical design for pair and joint pair/Compton telescopes is a low-Z tracker atop a high-Z calorimeter.

When developing a tracker for space-based gamma-ray science, a typical starting point is the semiconductor silicon (Si). Si is a historically popular detector material for trackers in space-based high-energy and astroparticle experiments, including the soft gamma-ray detector (SGD) aboard ASTRO-H \cite{astroh} and the tracker for the Fermi Large Area Telescope (LAT) \cite{fermi_lat}. Si is frequently chosen as a detector material partly because it is a prevalent technology that is relatively cheap to manufacture. Because particle tracking in three dimensions is necessary to reconstruct Compton events, the Si for a combined pair/Compton tracker has to record position information in three dimensions as well as convey the energy associated with interactions in the material. This constrains a joint pair/Compton telescope to two types of Si technologies: double-sided silicon strip detectors (DSSDs) \cite{griffin} or Si pixels. 

DSSDs provide precise position and energy information within a single layer of Si and are therefore a technology candidate for exploring soft- and medium-energy gamma-ray sky. However, there are a few drawbacks to DSSDs. DSSDs are traditionally expensive to manufacture because there are numerous mask layers required in the processing. They also require separate readout electronics. Therefore, to make a detector with large active area, multiple detectors have to be chained together with readout electronics located potentially far away. The chaining of DSSDs, combined with the fact that they are already long strips, means that DSSDs have high capacitance and hence a lot of noise. A high noise floor from the instrument interferes with the ability to finely resolve events at lower energies, impacting the low-energy range of an instrument. It is critical to reduce noise from the detectors in order to be fully sensitive to events closer to 100 keV. Monolithic Si pixels, a variation on traditional Si detectors, could provide an alternative solution to DSSDs.

A pixelated Si detector could provide the 3D position information within a single layer that is necessary for Compton reconstruction. Monolithic pixels combine the sensitive area of the detector with readout electronics, resulting in a co-integrated detector that can both detect and readout an event. Motivated by next-generation space-based high-energy missions, a NASA Astrophysics Research and Analysis (APRA) project, AstroPix, is working to adapt existing monolithic Si pixel technology for the detection of gamma-rays. The AstroPix project is interested in using Complementary Metal-Oxide-Semiconductor (CMOS) technology as a way to integrate readout electronics into silicon detectors, allowing us to gauge the potential of monolithic Si pixels in gamma-ray astrophysics \cite{spie}. The co-integration of detector and readout capabilities dispenses with the need for a separate application-specific integrated circuit (ASIC). In addition to saving on space, this means that monolithic Si pixels do not have to be chained together, which reduces the capacitance, and therefore the noise, of the detector. This reduction in noise improves the energy resolution, and in turn angular resolution, of the detector, a key parameter in the reconstruction of Compton events.

In this paper, we investigate the feasibility of using CMOS silicon pixel detectors in future gamma-ray space missions. Section \ref{sec:Sims} details simulations for AstroPix detector configurations. These simulations help us to form parameters for the eventual AstroPix pixels and shape our understanding of how the prototype AstroPix detector, ATLASPix, can be customized for space-based astrophysics. In Section \ref{sec:ATLASPix}, the ATLASPix detector design will be discussed, as well as lab procedure and detector characterization of ATLASPix. Section \ref{sec:AstroPix} outlines the overall AstroPix project, and in the Section \ref{sec:future}, the future of AstroPix and plans for upcoming generations of pixels are discussed.

\section{\large Simulations}
\label{sec:Sims}
\noindent 

Simulations were carried out in the Medium Energy Gamma-ray Astronomy Library (MEGAlib) \cite{megalib} to ascertain ideal pixel size and thickness of the detector for a full-scale telescope based on the AstroPix design.
\textcolor{black}{While it is straightforward to estimate the angular resolution in the pair regime, performance in the Compton regime is most easily determined through simulations.}

\begin{figure}
\centering
\includegraphics[width=0.6\textwidth]{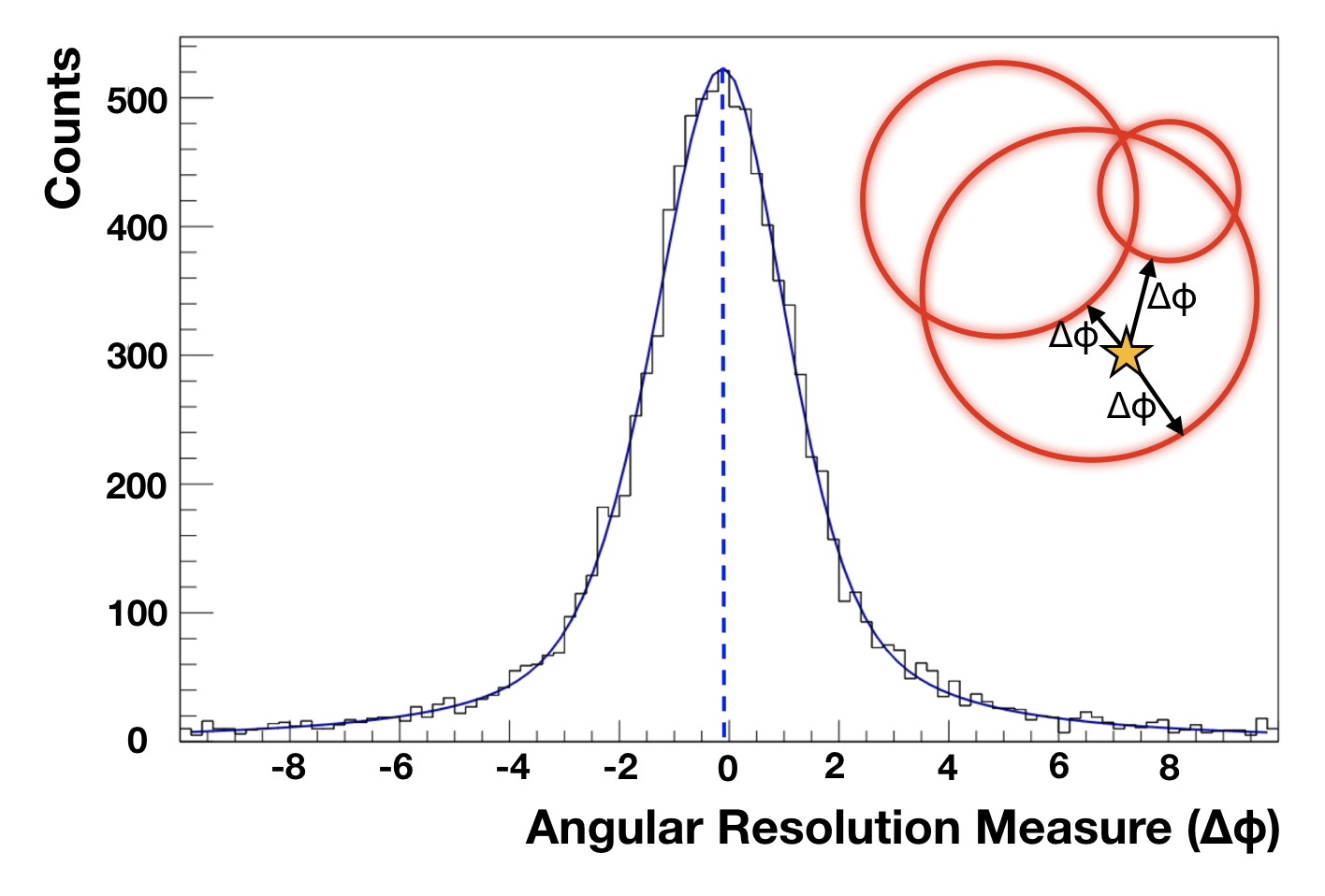}
\caption{\textcolor{black}{The Angular Resolution Measure (ARM) is a measurement of the shortest angular distance between the photon origin and the Compton event circle, as shown in the insert. The distribution of many ARM values indicates the accuracy of the Compton reconstruction, with the FWHM being the standard definition of the angular resolution in the Compton regime \cite{andreas}.}}
\label{fig:ARMexample}
\end{figure}

\textcolor{black}{Two of the main figures of merit for any imaging telescopes are the angular resolution and the effective area. 
In a Compton telescope, a precise measurement of the position and energy of interactions can constrain the origin of a photon to a circle on the sky, as shown in Figure~\ref{fig:eventschematic} and described in Sec.~\ref{sec:design}. The angular resolution is given by the FWHM of the distribution of the angular resolution measure (ARM) for each event, where the ARM is the smallest angular distance between the reconstructed event circle and the true position of the source. \textcolor{black}{An example of the ARM distribution is shown in Figure \ref{fig:ARMexample}, where the insert shows the defined the Angular Resolution Measure ($\Delta \phi$) for 3 event circles.}} 
The effective area is a measure of the efficiency of a telescope, and here we report results of the simulation as the number of events which were properly reconstructed and passed the event selections (see below) divided by the initial number of simulated events. 

We defined a baseline geometry of our instrument as composed of 50 layer tracker and ideal calorimeter to measure the energy and position of the Compton-scattered photon. 
This geometry is similar to the current AMEGO mission \cite{mcenery} \textcolor{black}{and is shown in Figure~\ref{fig:eventschematic}}. 
Each layer of the tracker is modelled as a 
1 $\times$ 1 m$^{2}$ sheet of pixelated silicon with 1 cm between layers. This detector model was placed in an enclosing volume made of vacuum.
Since the reconstruction of events in a Compton telescope depend strongly on the energy resolution and position resolution, we were interested to explore the effect of the pixel size. We tested square pixels from 0.01 $\times$ 0.01 mm$^2$ up to 10 $\times$ 10 mm$^2$. 
The thickness of the silicon layers is set to 500 $\mu$m for the results shown here
but we have simulated additional thicknesses to understand the instrument response
(100 $\mu$m and 700 $\mu$m). 
An ideal calorimeter is located underneath the tracker and is modelled as a 3-D position sensitive detector whose function is to provide an accurate measure of the Compton-scattered photon so that the calorimeter itself does not limit the telescope performance. 
For every configuration, we simulate a monochromatic far-field point source at four energy values: 200~keV, 300~keV, 500~keV and 1~MeV.


We adopt a classic Compton sequence event reconstruction algorithm available in MEGALib-revan library \cite{zoglauer2006}.
The \textcolor{black}{simulated} angular resolution is evaluated on a subset of selected events whose reconstructed energies fall under the peak of the distribution centered on the true energy of the simulated source.

\begin{figure}
    \centering
    \includegraphics[scale=0.3]{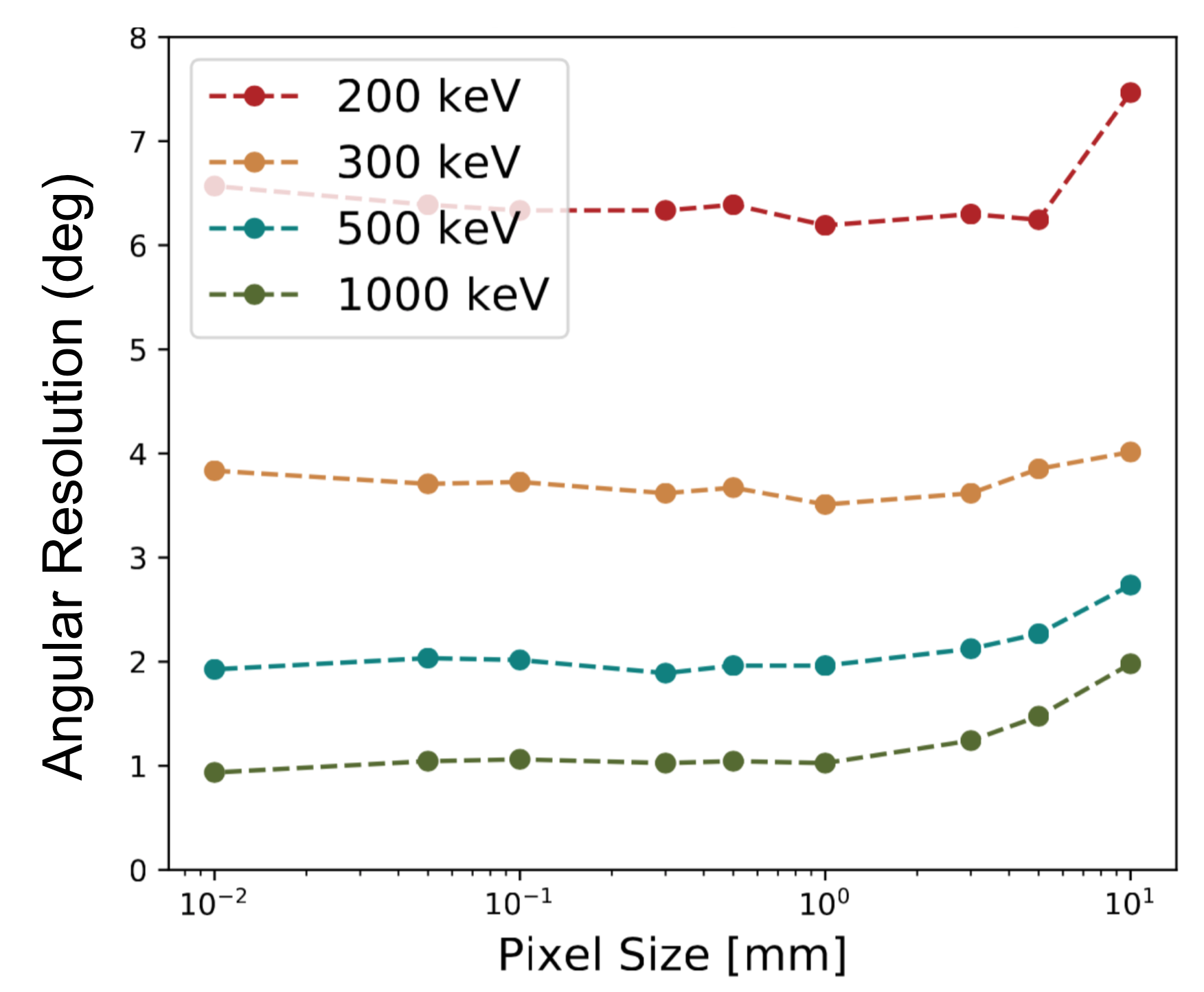}
    \caption{Angular resolution 
    as a function of the pixel size for different monochromatic energies for 500 $\mu$m thick pixels. The fraction of passive material is set to 5$\%$.}
    \label{fig:ARMenergy}
\end{figure}

Figure \ref{fig:ARMenergy} illustrates the angular resolution 
as a function of the pixel size for different monochromatic sources for 500 $\mu$m thick pixels. 
We observe the expected trend of the improvement of the \textcolor{black}{simulated} angular resolution with increasing energy. Also, for all the energies, we can appreciate a plateau in the angular resolution curve as a function of the pixel size up to $\sim$  1 $\times$ 1 mm$^2$. The uncertainty in both the measured position and energy contribute to the angular resolution, but at smaller pixel sizes the energy resolution dominates. This accounts for the plateau. This suggests we can have a pixel size of up to 1 $\times$ 1 mm$^2$ without impacting the angular resolution, which results in a lower number of readout channels and consequently in a lower amount of total power consumption (which is crucial aspect for any space-based instrument).

\begin{figure}[!htbp]
    \centering
    \begin{subfigure}[!htbp]{0.473\textwidth}
        \centering
        \includegraphics[width=\textwidth]{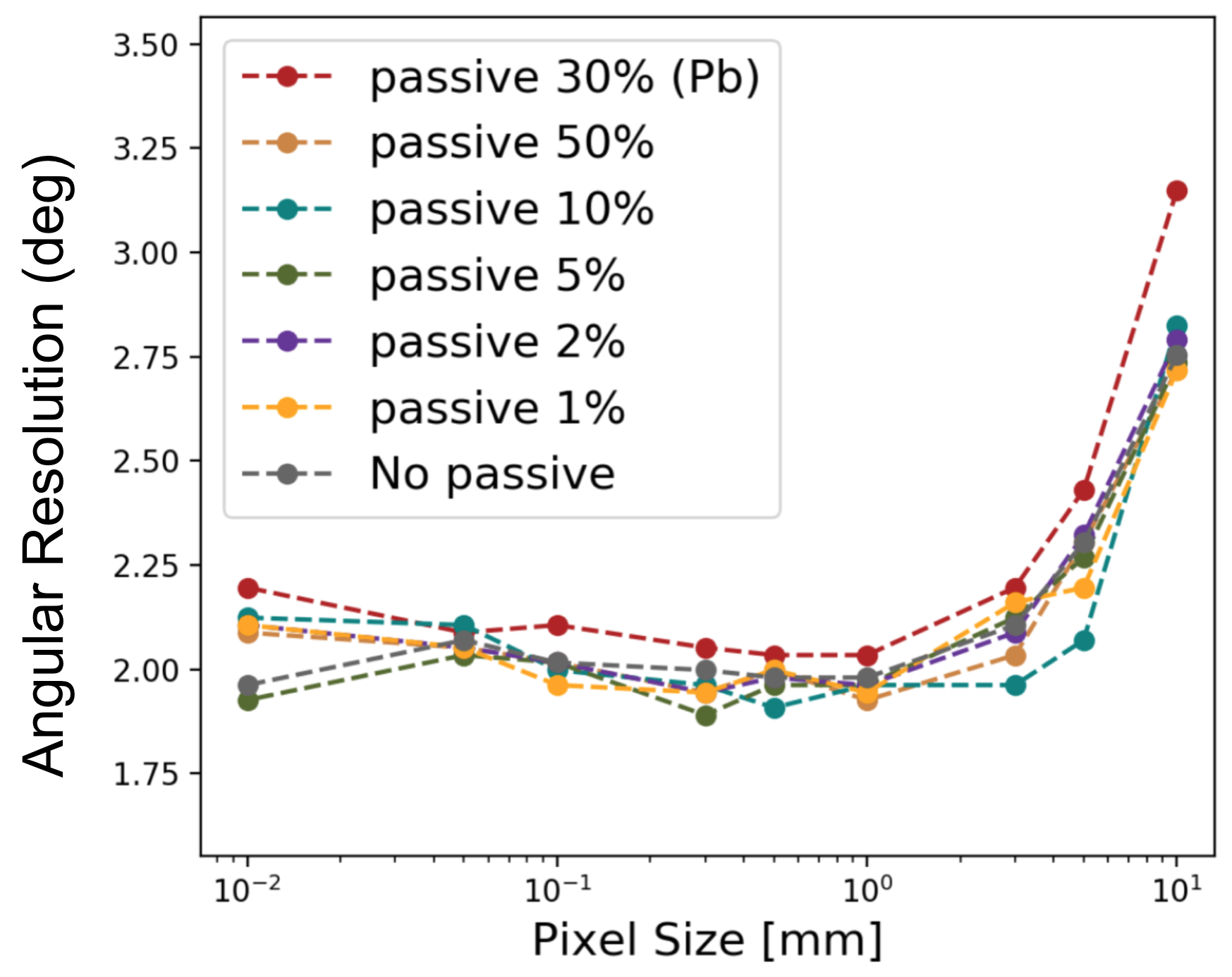}
        \caption{}
        \label{ARMpassive1}
    \end{subfigure}
\hfill
    \begin{subfigure}[!htbp]{0.507\textwidth}
        \centering
        \includegraphics[width=\textwidth]{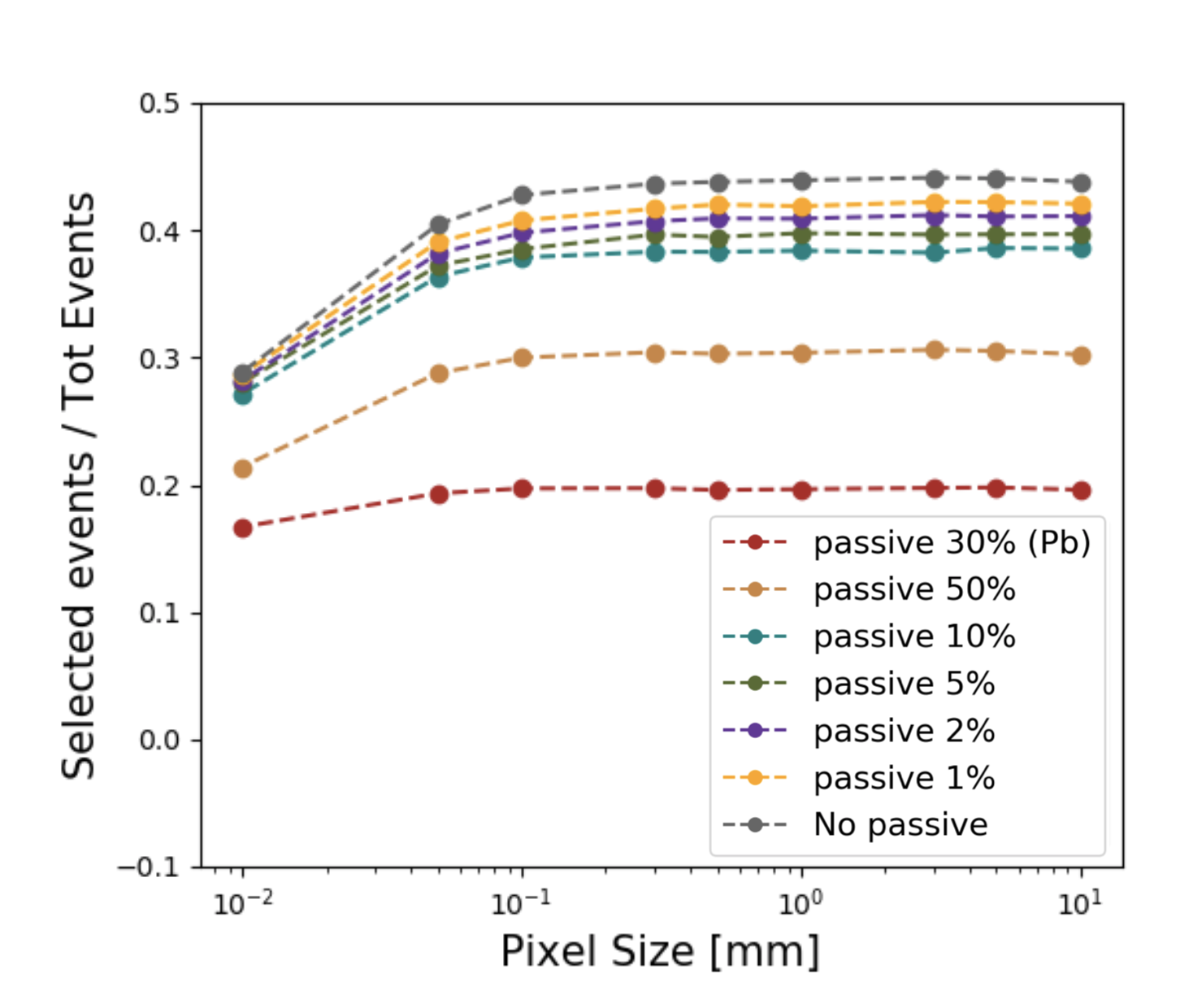}
        \caption{}
        \label{ARMpassive2}
    \end{subfigure}
    \caption{Study of the effect of different fractions (and kind) of passive material in the tracker at 500 keV. Note that in both plots the 30$\%$ case uses lead as passive material. (a) Angular resolution as a function of the pixel size. (b) Fraction of events passing the selection cuts as a function of the pixel size.}
    \label{fig:ARMpassive}
\end{figure}

\textcolor{black}{We studied the effect of passive material by applying additional non-sensitive silicon material at 0.1 cm beneath each sensitive layer. 
The thickness of the passive layers varies so that the total amount of passive material represents a given percentage of the total active tracker mass.} Figure \ref{fig:ARMpassive} shows the effect of the passive material on the angular resolution 
and the fraction of events which pass the energy cuts. For this study we focus on a median energy value in the Compton range, 500 keV. As we can observe from \textcolor{black}{Figure \ref{ARMpassive1}}, the amount of passive material does not affect the angular resolution estimation (only when we simulate 30\% of lead between the detector layers we start to appreciate a slight worsening of the angular resolution at every pixel sizes). The presence of the passive material dramatically affects the efficiency of the event selection, \textcolor{black}{as seen in Figure \ref{ARMpassive2}}, and hence the statistics we can accumulate.

Based on the angular resolution values seen in Figure \ref{ARMpassive1}, a pixel size less than 1 mm$^2$ is desirable. Also, we observe that the efficiency starts to systematically decrease when the pixel size becomes lower than about 0.1 - 0.3 mm$^2$: this is the effect of the smaller pixel volumes, when less energy is deposited in a single pixel, while the energy threshold to trigger the pixel is kept fixed to 20 keV in our simulation setup. Therefore, we are left with a range of optimal pixel sizes between 0.1 - 1 mm$^2$, which given systematic and statistical errors, behave about the same. The power consumption of the chip drives the choice of large pixel sizes, while the requirement to fully deplete each pixel is more easily achieved with smaller pixel sizes. Balancing these factors led to the selection of the 500 x 500 $\mu$m pixel size for AstroPix.

\section{\large ATLASPix: A Prototype for Next-Generation Gamma-ray Missions}
\label{sec:ATLASPix}
\noindent

The AstroPix project seeks to optimize monolithic Si pixels for tracking detectors in high-energy astroparticle experiments. The design for AstroPix is based on the simulations discussed in the previous section and on an existing high voltage CMOS (HVCMOS) detector called ATLASPix. ATLASPix was developed for the ATLAS experiment \cite{atlas} at CERN (European Organization for Nuclear Research) and optimized for the detection of high-energy particles, specifically minimum ionizing particles (MIPs). The ATLASPix optimization for MIPs drives the design to have limited spectral resolution and very accurate timing resolution for thin, radiation-hard detectors. To determine the design parameters for AstroPix pixels, the behavior of ATLASPix when exposed to high energy photons had to be understood. The performance of an existing monolithic Si detector, although not optimized for the detection of gamma-rays, will indicate the suitability of monolithic pixels for space-based astrophysics research and will also highlight ways in which ATLASPix can be re-customized. 

\begin{figure}[!htbp]
    \centering
    \includegraphics[width=.7\textwidth]{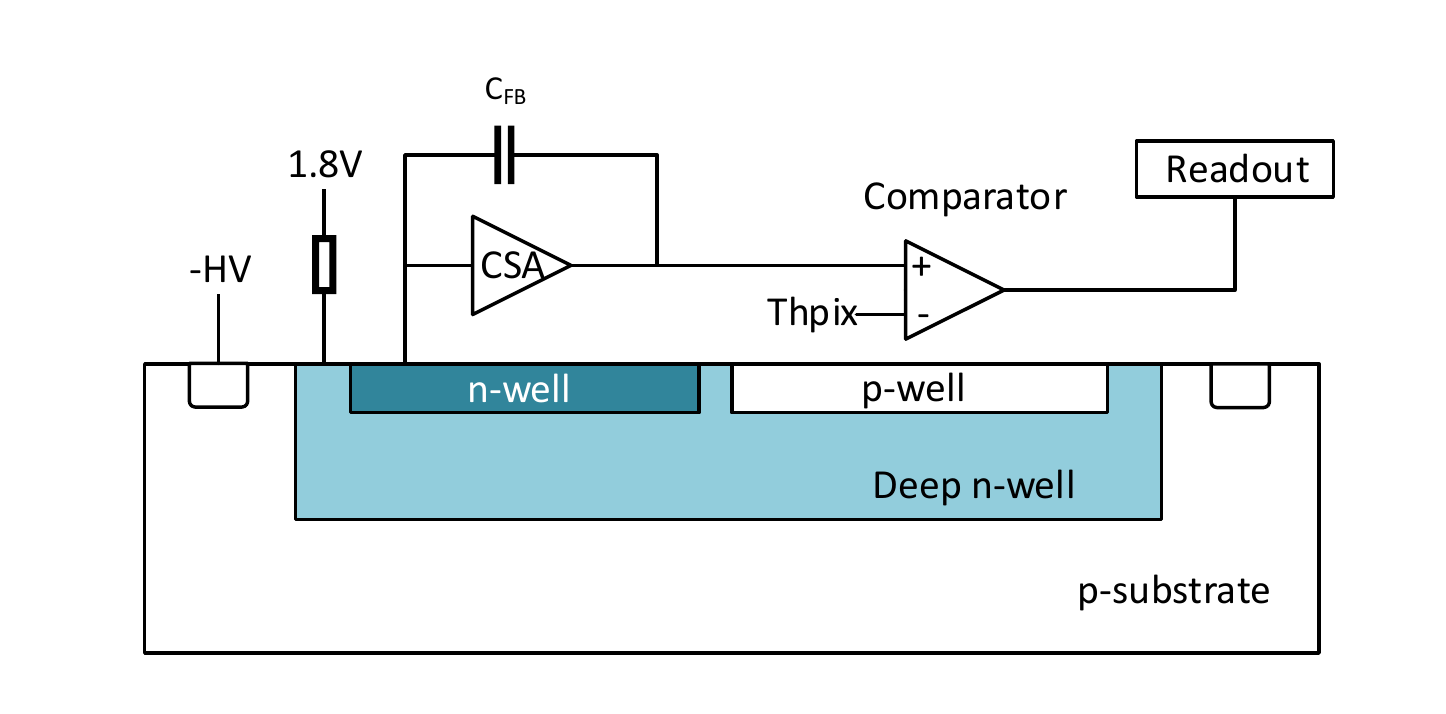}
    \caption{Diagram of an ATLASPix-type pixel \cite{nicolas}. The detection and readout capabilities are co-integrated into a single pixel, reducing the noise of the readout. The size of the deep n-well, shallow n-well, and p-well are exaggerated for clarity; in reality, the embedded electronics take up much less of the sensitive area of the pixel.}
    \label{nicolas_diagram}
\end{figure}

ATLASPix pixels have a shallow n-well embedded within a p-substrate, as seen in Figure \ref{nicolas_diagram}. The shallow n-well contains the readout electronics (circuits and logic gates) and amplification. This co-integration of readout electronics with the sensitive area of the detector makes it possible to amplify the signal right at the source, reducing the noise of the readout. The isolation of the n-well from the p-substrate permits a high bias voltage to be applied to the substrate, allowing for the detector volume to be fully depleted \cite{ivan_1}. The ATLASPix$\_$Simple matrix was used for these initial characterization tests \cite{schoning}; this detector version consists of four matrices of 25 by 100 pixels, with each matrix stacked vertically and chained together to create an active area of 25 by 400 pixels. Each pixel is 130 $\mu$m $\times$ 40 $\mu$m and 100 $\mu$m thick.  

\subsection{\large \textit{Experimental Setup}}
\label{sec:Setup}
\noindent

\begin{table}[ht]
\caption{Sources used to illuminate the ATLASPix detector along with the relevant line energy \cite{pdg} \cite{TabRad_v1}. Three sources (Fe55, Cd109, and Ba133) were sealed laboratory radioactive disk sources and three sources (Ge, Y, and Mo) were created by having accelerated electrons hit different source targets. These energies were selected because they span the dynamic range of ATLASPix.} 
\label{tab:fonts}
\begin{center}       
\begin{tabular}{|c|c|}
\hline
\rule[-1ex]{0pt}{3.5ex}  \textbf{Source} & \textbf{Energy (keV)}  \\
\hline
\rule[-1ex]{0pt}{3.5ex}  Fe55 & 5.89  \\
\hline
\rule[-1ex]{0pt}{3.5ex}  Ge & 9.89  \\
\hline
\rule[-1ex]{0pt}{3.5ex}  Y & 14.96  \\
\hline
\rule[-1ex]{0pt}{3.5ex}  Mo &  17.5 \\
\hline
\rule[-1ex]{0pt}{3.5ex}  Cd109 &  21.99  \\
\hline
\rule[-1ex]{0pt}{3.5ex}  Ba133 &  30.97 \\
\hline
\end{tabular}
\end{center}
\label{source_table}
\end{table}

The first step of AstroPix development required characterizing the behavior of the ATLASPix detector when exposed to high energy photons. Six different radioactive sources were used gauge the energy response of ATLASPix, listed in Table \ref{source_table}. Three of the sources-- Fe55, Cd109, and Ba133-- were isotropic radioactive disk sources that were placed in front of the detector. The other three sources-- Ge, Y, and Mo-- were created by hitting target materials with accelerated electrons to create a collimated monolithic x-ray source. Ideally for an MeV Compton telescope, a single pixel would be sensitive up to $\sim$700 keV. However, ATLASPix is comprised of a single layer of Si pixels 100 $\mu$m thick, meaning that the chance of higher energy photons interacting in the material is negligible. As a result, lower energy sources were used for calibration that spanned the dynamic range of the detector from $\sim$4 to 35 keV.

\begin{figure}
        \centering
        \includegraphics[width=.6\textwidth]{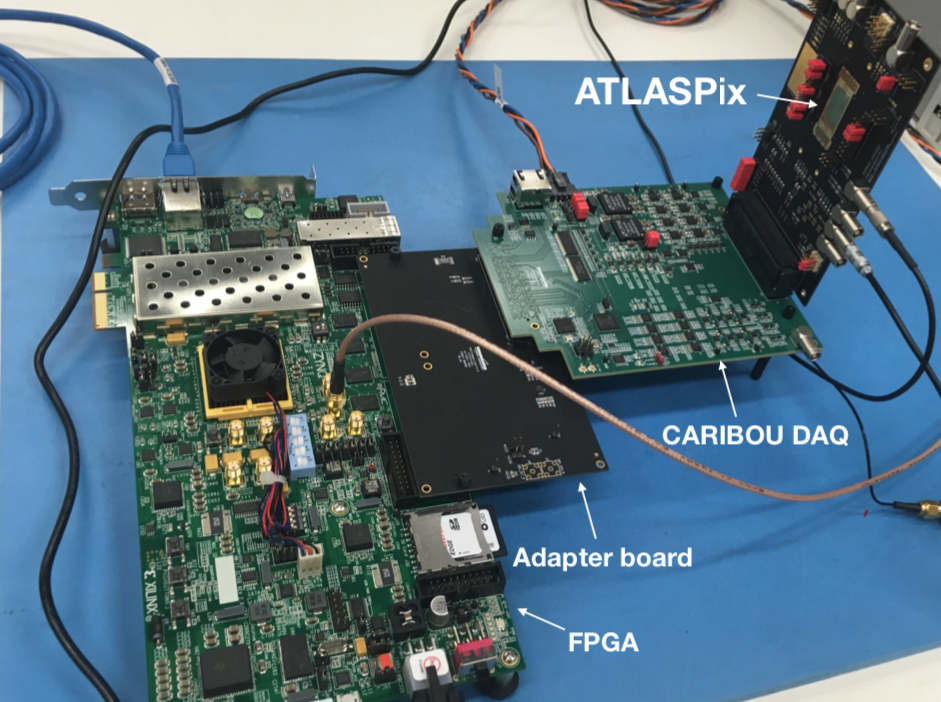}
        \caption{Test setup for ATLASPix. This includes (from left to right) a Zync FPGA that contains software to communicate to the device, an adapter board, a custom CaRIBOu DAQ for processing data from ATLASPix, and the ATLASPix detector situated vertically on a carrier board.}
        \label{test_setup}
\end{figure}

The equipment used to test ATLASPix, as seen in Figure \ref{test_setup}, included a custom built Control and Readout Inner tracking Board (CaRIBOu) Data Acquisition (DAQ) System made by Brookhaven National Laboratory, a Zynq ZC706 Field Programmable Gate Array (FPGA), and an oscilloscope for digitization. All of the sealed disk sources (Fe55, Cd109, Ba133) were placed approximately 2 cm in front of the detector, while the aperture for the x-ray sources was placed approximately 20 cm away from the detector.

\begin{figure}[!htbp]
    \centering
     \begin{subfigure}[!htbp]{0.48\textwidth}
        \centering
         \includegraphics[width=\textwidth]{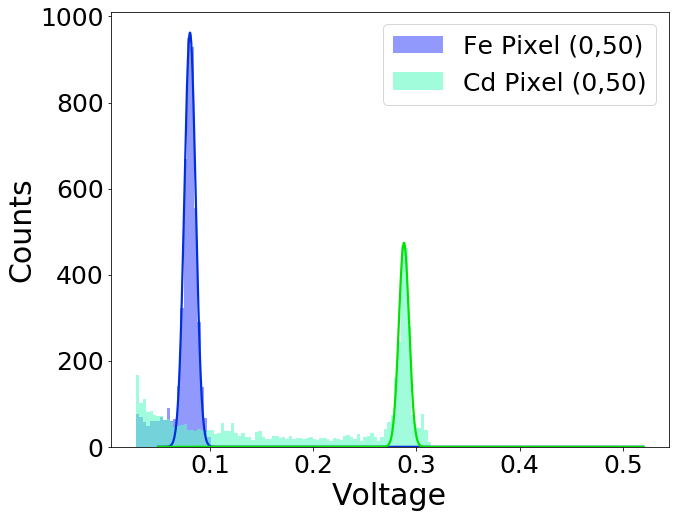}
         \caption{Pixel (0,50).}
         \label{comparing_photopeak_positions_0_50}
     \end{subfigure}
     \hspace{0.005\textwidth}
     \hfill
     \begin{subfigure}[!htbp]{0.48\textwidth}
         \centering
         \includegraphics[width=\textwidth]{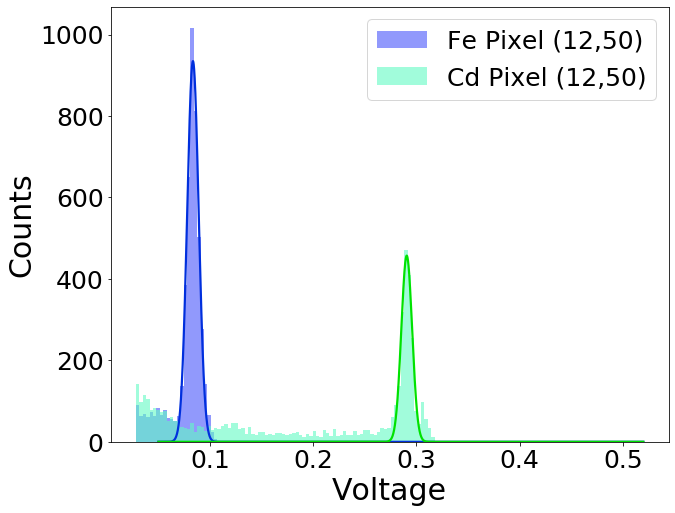}
         \caption{Pixel (12,50).}
         \label{comparing_photopeak_positions_12_50}
     \end{subfigure}
     \caption{Plots of the Fe55 (peaks at 5.89 keV) and Cd109 (peaks at 22.1 keV) photopeaks as detected by two different pixels. The data was collected using the analog output and is uncalibrated. Comparing the centroid positions of the Fe55 and Cd09 photopeaks between pixels, it was found that the centroid positions agree within errors, which suggests some uniformity in pixel response. It took significantly longer to collect data for the Fe55 photopeaks, Fe55 being a colder source.}
     \label{fig:comparing_photopeak_positions}
\end{figure}

ATLASPix has both an analog and digital output. The analog output of ATLASPix, which is the output of the comparator in the pixel prior to the standard digitization readout path, is a voltage output that can only be read from a single pixel at a time. Importantly, the analog output corresponds to the intrinsic resolution of the detector and amplifying electronics and is not limited by the digital resolution of the readout. The analog output of ATLASPix was connected to the oscilloscope. For each source, 5000 digitized waveforms were collected from a single pixel; the pedestal value was recorded and a noise sample (the mean of a measurement taken without the source present) was subtracted off the maximum voltage measurement. The pedestal-subtracted voltages were then histogrammed to plot the energy spectrum of each source.

Uncalibrated photopeaks from the analog output can be seen in Figure \ref{fig:comparing_photopeak_positions}. Two pixels within the same matrix were compared to see how photopeak position changed between pixels. The same two sources, Fe55 and Cd109, were exposed at different times to pixels (0,50) and (12,50) (where pixel position is given by (x,y)). Figure \ref{fig:comparing_photopeak_positions} corresponds to two data sets taken from a single pixel, with (0,50) \textcolor{black}{shown in Figure \ref{comparing_photopeak_positions_0_50}} and (12,50) \textcolor{black}{shown in Figure \ref{comparing_photopeak_positions_12_50}}. Comparing the centroids of the Fe55 photopeaks and the centroids of the Cd109 photopeaks, the values agree within errors. It suggests uniformity in pixel response, although further measurements would have to be carried out to confirm this. 

The digitization occurs outside of the pixel in the periphery of the chip. The digital output is processed by the DAQ to produce a text file that contains, for each hit incident with a pixel, the pixel number and approximate energy deposited in time over threshold (TOT). The digital output has \textcolor{black}{a timing resolution of 12.5 ns.} This is consistent with the timing resolution necessary for the ATLAS experiment. The digital output's proxy for the energy deposited in the pixel, TOT, has a 6 bit resolution, which is sufficient for MIPs but too low of a resolution for the reconstruction of Compton scatter events. 

A process called tuning can be applied to the ATLASPix pixels. When tuning, a value stored in the RAM can be configured to change the local comparator threshold of each individual pixel, making the detection threshold, or the value at which the trigger threshold results in 50\% probability of detection, as small as possible. Optimizing the detection threshold for each pixel results in improved detector performance \cite{ivan_1}. Tuning is possible for MIP detections where the energy deposits are known, but the \textcolor{black}{energy range} of photon sources that ATLASPix used meant that the tuning process was not effective for our measurements. 

\subsection{\textit{Energy Calibration of ATLASPix}}
\noindent

It was necessary to calibrate and assess ATLASPix using both its readout methods, digital and analog. Understanding detector response using the digital output is important because it reveals how the digital processing power was assigned and how these capabilities can be redesigned. Likewise, assessing ATLASPix performance using the analog output is critical because the analog output is a better representation of the achievable spectral resolution.

\begin{figure}[!htbp]
    \centering
     \begin{subfigure}[!htbp]{0.5\textwidth}
         \centering
         \includegraphics[width=\textwidth]{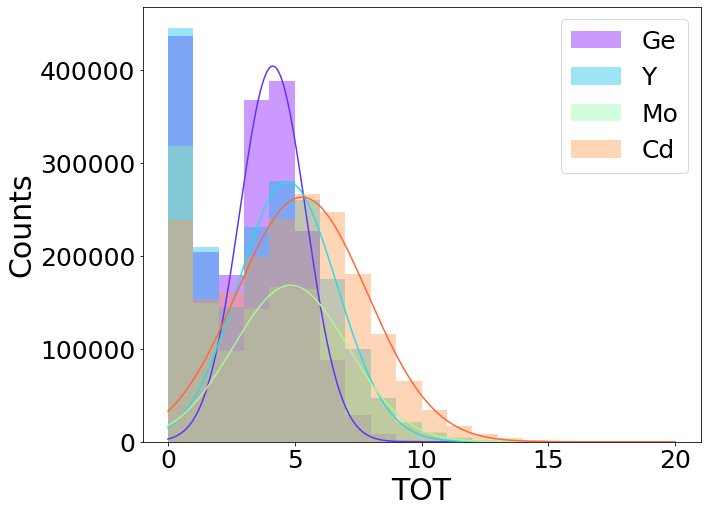}
         \caption{}
         \label{daq_photopeaks}
     \end{subfigure}
     \begin{subfigure}[!htbp]{0.465\textwidth}
         \centering
         \includegraphics[width=\textwidth]{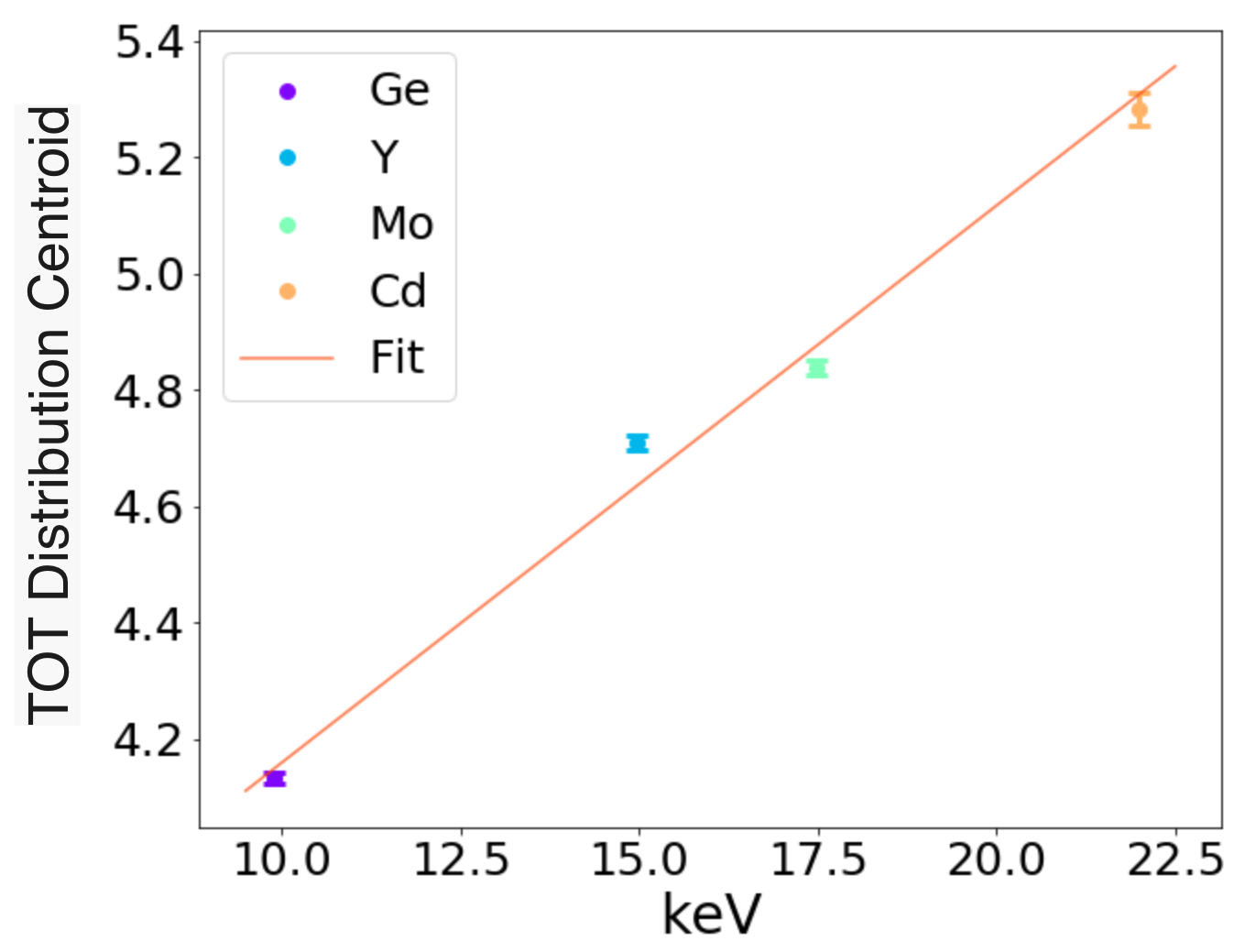}
         \caption{}
         \label{daq_energy_calibration}
     \end{subfigure}
     \caption{(a) Four photopeaks using the TOT values extracted from the DAQ files. Each uncalibrated photopeak from the digital output in was separately fit with a Gaussian distribution, and the resulting FWHM returned $\Delta E / E$ resolution of greater than 100$\%$. (b) Energy calibration using the digital output. The error bars, \textcolor{black}{representing the statistical error found when calculating the position of the photopeak}, resulted in a one degree polynomial being sufficient to represent the detector response.}
     \label{daq_energy_scaling}
\end{figure}

To create a spectra using the digital output, \textcolor{black}{t}he detector was exposed to the sources from Table \ref{source_table}. For each source, the TOT values from the digital output were histogrammed to plot the photopeaks. A Gaussian distribution was fit to each photopeak in \textcolor{black}{Figure \ref{daq_photopeaks}} to determine the centroid of each peak and the width $\sigma$.

Characterizing the detector response using the digital output was achieved by comparing the measured photopeak positions to the known energies of the emission lines. The photopeaks of four sources using the digital output can be seen in \textcolor{black}{Figure \ref{daq_photopeaks}}. A noise peak occupies the first bin of the plot. The TOT values from all pixels are combined to a single histogram. The TOT values extracted from a single pixel provided too few statistics for the calibration of the digital output, and demonstrated the same trend of an extremely broad resolution, resulting in our choice to use TOT values from across the detector in order to perform the calibration. It should be noted that Figure \ref{daq_photopeaks} should not be compared directly with Figure \ref{analog_resolution}, as Figure \ref{analog_resolution} uses a single pixel for calibration. The \textcolor{black}{energy scaling calibration} for the digital output, which compares the energy of each source  in keV to centroid of each photopeak in TOT, can be seen in \textcolor{black}{Figure \ref{daq_energy_calibration}}. \textcolor{black}{Due to the size of the error bars, representing the statistical error in calculating the centroid of the photopeaks,} a first order polynomial was found to be sufficient to \textcolor{black}{describe the  detector linear response} \textcolor{black}{depicted in Figure \ref{daq_energy_calibration}}. 

Using the full-width half-max (FWHM) to calculate the energy resolution, each photopeak had a $\Delta E / E$ resolution of greater than 100$\%$. This is expected, as ATLASPix was optimized for MIPs and not for photons, and the energy proxy from the digital output-- a six bit TOT value-- constrained the energy resolution.

To characterize ATLASPix response using analog data, the sources from Table \ref{source_table} were exposed, at separate times, to the detector. The data collection and analysis method for constructing analog photopeaks involved using the oscilliscope to record digital waveforms and peak heights, the same method used to create Figure \ref{fig:comparing_photopeak_positions}. After photopeaks had been plotted for each source, the source energies in keV was compared to the central photopeak positions in volts in order to determine the detector response. 


\begin{figure}[!htbp]
    \centering
        \includegraphics[width=.65\textwidth]{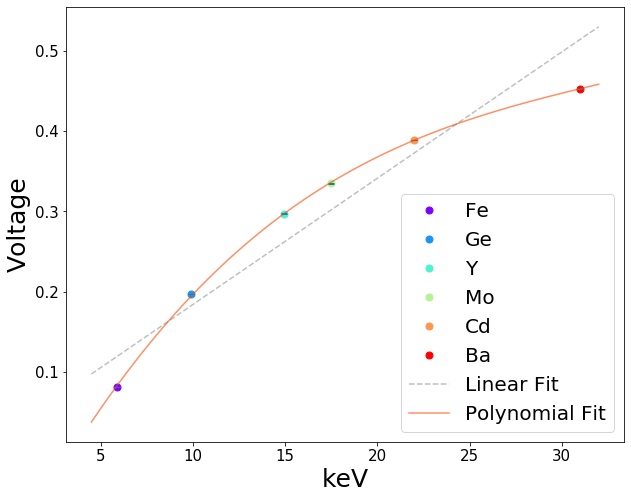}
        \caption{Energy calibration showing the detector response using the analog output. A single pixel (pixel (0,50)) was used for calibration. A three degree polynomial was found to best represent the detector response, as it allowed for energy reconstruction of events within 1\% of the reference values. The error bars represent the \textcolor{black}{statistical error in calculating the position of the centroid of each photopeak}.}
        \label{analog_scaling}
    \end{figure}
    
Figure \ref{analog_scaling} shows the calibration between keV and volts; a linear detector response (shown in grey) is compared to a three degree polynomial response. The error bars represent \textcolor{black}{the statistical error when calculating the position of the photopeaks}. While determining the fit that best represents the detector response, a two degree polynomial fit had a reduced $\chi^2$ of 1.1 and was found statistically to be sufficient for our data. However, a two degree polynomial fit only allowed us to reconstruct energies within 5\% of the reference values. A three degree polynomial allowed for reconstruction within 1\% of the reference values, therefore best representing the detector response. A linear detector response would be optimal, but it is expected that the detector and amplifier would have a nonlinear response. After the detector response had been ascertained, it was possible to apply this to the analog data in order to find the calibrated energy resolution of the detector.

\subsection{\textit{Analog Energy Resolution of ATLASPix}}
\noindent

\begin{figure}[!htbp]
    \centering
        \includegraphics[width=.65\textwidth]{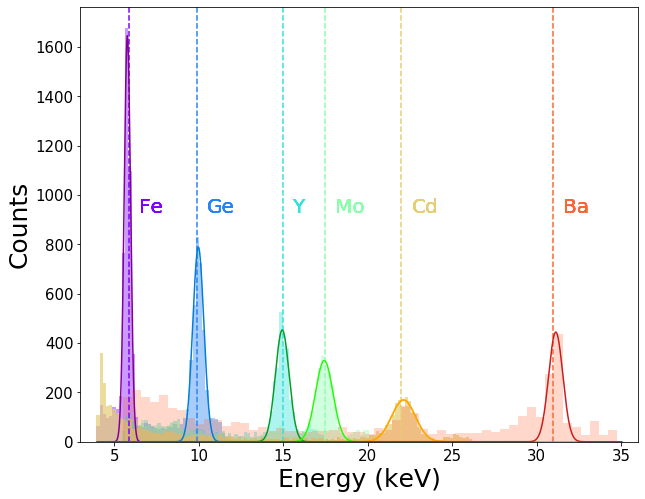}
        \caption{The analog energy resolution of the ATLASPix detector using a single pixel (pixel (0,50)). The calibrated photopeaks represent the six sources in Table \ref{source_table}; the resolution at each source energy exceeds the minimum goal of $<$10\%.} 
        \label{analog_resolution}
    \end{figure}

\begin{figure}[!htbp]
    \centering
        \includegraphics[width=.6\textwidth]{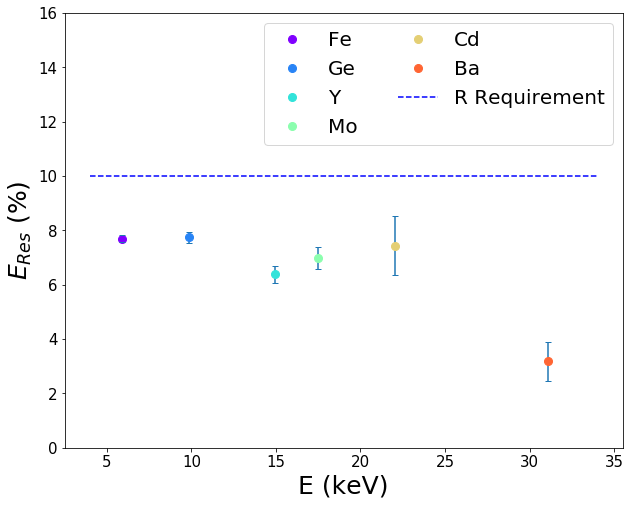}
        \caption{The energy resolution of pixel (0,50) plotted at each source energy. The horizontal dashed line represents the minimum resolution requirements of AstroPix, and the error bars are the 1 $\sigma$ statistical error. \textcolor{black}{We would expect to see a 1/$\sqrt{E}$ relationship between the E$_{Res}$ and the energy; however, the non-linear detector response distorts this relationship, possible indicating that the resolution is not dominated by the stochastic processes of the detector.}}
        \label{e_v_r}
    \end{figure}

Energy resolution is a key metric for Compton event reconstruction in gamma-ray astrophysics, and as such it is a driving parameter for the AstroPix project. Because of the importance of fine energy resolution, the analog energy resolution, which represents the intrinsic resolution of the detector and amplifier, is a crucial measurement. Six analog photopeaks were fit with a Gaussian distribution and, with the scaling from Figure \ref{analog_scaling} applied, are shown in Figure \ref{analog_resolution}. The vertical dashed lines represent the known energy of the emission lines from each source in keV. Figure \ref{e_v_r}, which depicts the analog resolution as a function of energy, shows the baseline requirement of 10\% as a horizontal dashed blue line. The baseline requirement of $<$10\% is exceeded at each source energy, even including error bars (which represent the statistical error found in each energy resolution calculation). \textcolor{black}{There should be a 1/$\sqrt{E}$ relationship between the energy and the resolution, which is not seen in Figure \ref{e_v_r}. This discrepancy is likely the result of the nonlinear response of the detector, which possibly indicates that the resolution is not dominated by the stochastic processes of the detector. }

Using the FWHM from the Gaussian distribution fit, the energy resolution was found to be 7.7 $\pm$ 0.1\% at 5.89 keV and \textcolor{black}{3.18 $\pm$ 0.73 \%} at 30.1 keV. For future versions of AstroPix with a thicker detector we expect the dynamic range to be increased. 
    
The digital energy resolution, depicted in \textcolor{black}{Figure \ref{daq_photopeaks}}, is not sufficient for reconstructing Compton events. However, the resolution of the digital output can be reconfigured for future generations of pixels and is a capability that can be redesigned. Comparing the digital and analog energy resolutions reveals a drastic difference in detector performance, pointing to the potential of re-optimized monolithic Si pixels for the detection of high energy photons.

\subsection{\textit{Hit Position and Clustering}}
\noindent 

\begin{figure}[!htb]
    \centering
     \begin{subfigure}[!htbp]{0.315\textwidth}
         \centering
         \includegraphics[width=\textwidth]{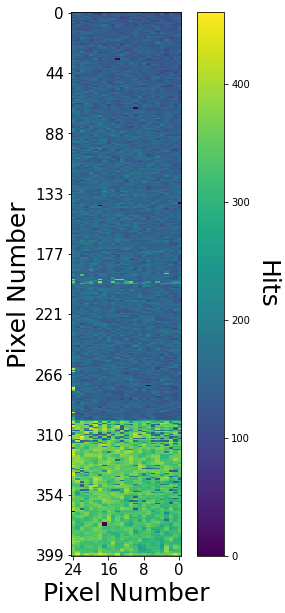}
         \caption{Hitmap for Cd109. The noisy bottom matrix of the detector dominates at the higher energies of Cd109. }
         \label{hitmaps_Cd109}
     \end{subfigure}
     \hspace{0.05\textwidth}
     \begin{subfigure}[!htbp]{0.325\textwidth}
         \centering
         \includegraphics[width=\textwidth]{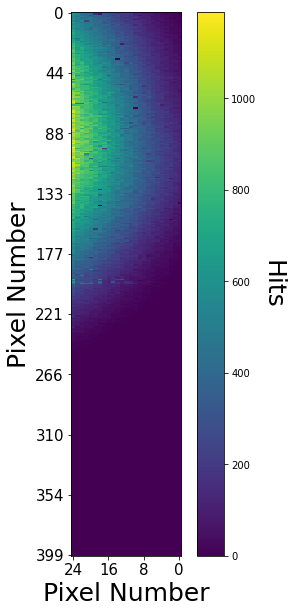}
         \caption{Hitmap for Fe55. The source can clearly be seen at the lower energy Fe55 source.}
         \label{hitmaps_Fe55}
     \end{subfigure}
     \caption{Hit position was extracted from the digital DAQ files taken for two different sources and plotted to show hit distribution and potential clustering.}
     \label{hitmaps}
\end{figure}

Charge sharing and cross-talk are two detector behaviors that could impact the characterization of ATLASPix. Charge sharing refers to the effect of the electron-hole cloud from a single hit drifting to multiple pixels, causing a loss of energy information. Cross-talk occurs when one electronics channel influences a neighboring channel, distorting or amplifying the output of the pixels. Searching for clusters of events in the digital data is one way to check for these detector behaviours. To investigate hit position and potential clustering in the detector, two sources, Fe55 and Cd109, were exposed at separate times to the detector and the digital output was recorded. The hit positions from the DAQ files were extracted and plotted to create hitmaps as seen in Figure \ref{hitmaps}. In \textcolor{black}{Figure \ref{hitmaps_Cd109}}, representing Cd109, the bottom matrix exhibited a much higher hit count, possibly revealing a noisy lower matrix. This pattern of hits \textcolor{black}{produced by the} Cd109 \textcolor{black}{source is} partly due to the logic chaining of the detector. In \textcolor{black}{Figure \ref{hitmaps_Cd109}}, the Fe55 source can be clearly seen. 

To search quantitatively for clustering, a depth-first search of the DAQ files was performed; this comprised of iterating over every hit and finding any pixels hit within an eight-pixel ring around the original hit and within $\pm$ 2 bins of 25 ns. If such a hit was detected, the search repeated with the new hit as the \enquote{pixel of interest,} searching for more pixels meeting the cluster criteria in a new ring around the pixel of interest. This process was repeated until the entire cluster size was ascertained, and then the search continued until a new cluster was detected. 1000 hits were parsed from each digital data set, and the results are shown in Table \ref{hitmap_table}.

\textcolor{black}{The small number of clusters detected for the Fe55 source indicates limited charge sharing at low energies, but unfortunately for Cd109 at 22~keV, 10\% of the events show clustering. The next generation of AstroPix pixels will also be larger than the current ATLASPix pixels, potentially reducing charge sharing. However, it is important to continue monitoring charge sharing as the project moves into higher energies. The broad time bin of 25 ns was chosen in order to investigate a conservative scenario. The low hit rate of the sources used meant we did not expect to see many coincident hits. }

\begin{table}[ht]
\caption{Results from the clustering search performed on the digital output data from across the detector. 1000 hits were parsed from each source data set. The number of clusters and average cluster size (in pixels) are shown. Clusters in one dimension are more likely than clusters in another direction because of the oblong shape of the pixels. } 
\label{tab:fonts}
\begin{center}       
\begin{tabular}{|c|c|c|}
\hline
\rule[-1ex]{0pt}{3.5ex}  \textbf{Source} & \textbf{Number of Clusters} & \textbf{Average Cluster Size} \\
\hline
\rule[-1ex]{0pt}{3.5ex}  Cd109 & 182 &  2.75 pixels \\
\hline
\rule[-1ex]{0pt}{3.5ex}  Fe55 & 11 & 2 pixels \\
\hline
\end{tabular}
\end{center}
\label{hitmap_table}
\end{table}

\section{\large AstroPix: Monolithic Silicon Pixels for Multimessenger Astrophysics}
\label{sec:AstroPix}
\noindent

To properly support high-energy multimessenger missions such as AMEGO-X, AstroPix needs to achieve baseline performance requirements in a few key areas, seen in Table \ref{goals_table}, including effective area, angular resolution, and energy resolution. The baseline energy resolution requirement for AstroPix \textcolor{black}{, based on the noise floor of 5 keV,} is \textcolor{black}{$\Delta E/E=10$\% FWHM at 60 keV}. Simulations (detailed in Section \ref{sec:Sims}) of AstroPix-style detectors suggest an optimal pixel size is 500 $\times$ 500 $\mu$m$^2$. AstroPix's goal for power consumption is about 1 mW/cm$^2$.

\begin{table*}[!ht]
\caption{Performance goals for AstroPix pixels. The first three goals are driving parameters of the project.} 
\begin{center}       
\begin{tabular}{|c|c|}
\hline
\rule[-1ex]{0pt}{3.5ex}  \textbf{Parameter} & \textbf{Goal}   \\
\hline
\rule[-1ex]{0pt}{3.5ex}  \textbf{E$_{Res}$} &  $<$10\% at \textcolor{black}{60} keV \\
\hline
\rule[-1ex]{0pt}{3.5ex}  \textbf{Power Usage} & $<$1 mW/cm$^2$   \\
\hline
\rule[-1ex]{0pt}{3.5ex} \textbf{Passive Material} &  $<$5\% on the active area of Si \\
\hline
\rule[-1ex]{0pt}{3.5ex}  Pixel Size &  500 $\times$ 500 $\mu$m$^2$  \\
\hline
\rule[-1ex]{0pt}{3.5ex}  Si Thickness &  500 $\mu$m \\
\hline
\rule[-1ex]{0pt}{3.5ex}  Time Tag &  $\sim$1 $\mu$s  \\
\hline
\end{tabular}
\end{center}
\label{goals_table}
\end{table*}

Pixels have a number of possible advantages compared to DSSDs, as discussed in Section \ref{introduction}, and the potential of this technology for the detection of high-energy photons is the focus of the AstroPix project. The objective of AstroPix is to produce multiple generations of AstroPix pixels that are progressively optimized for space-based multimessenger astronomy, with this path of optimization beginning with the characterization of an existing HVCMOS detector outlined in Section \ref{sec:ATLASPix}. 



\section{\large Outlook to Future AstroPix Generations}
\label{sec:future}
\noindent

\begin{figure}[!htbp]
    \centering
    \begin{subfigure}[!htbp]{0.43\textwidth}
        \centering
        \includegraphics[width=\textwidth]{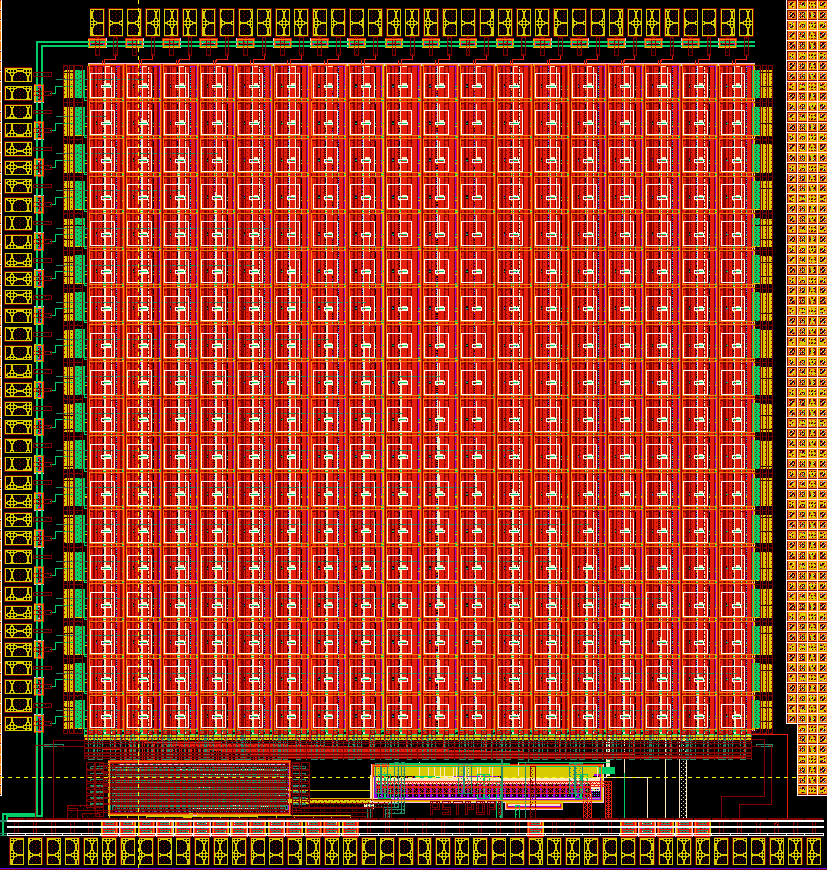}
        \caption{Top level view of the AstroPix V1 sensor matrix. The matrix is 18 by 18 pixels, each pixel 200 $\mu$m$^2$, with digital readout for each row and column.}
        \label{astropix_v1_diagram}
    \end{subfigure}
    \hfill
        \begin{subfigure}[!htbp]{0.47\textwidth}
        \includegraphics[width=\textwidth]{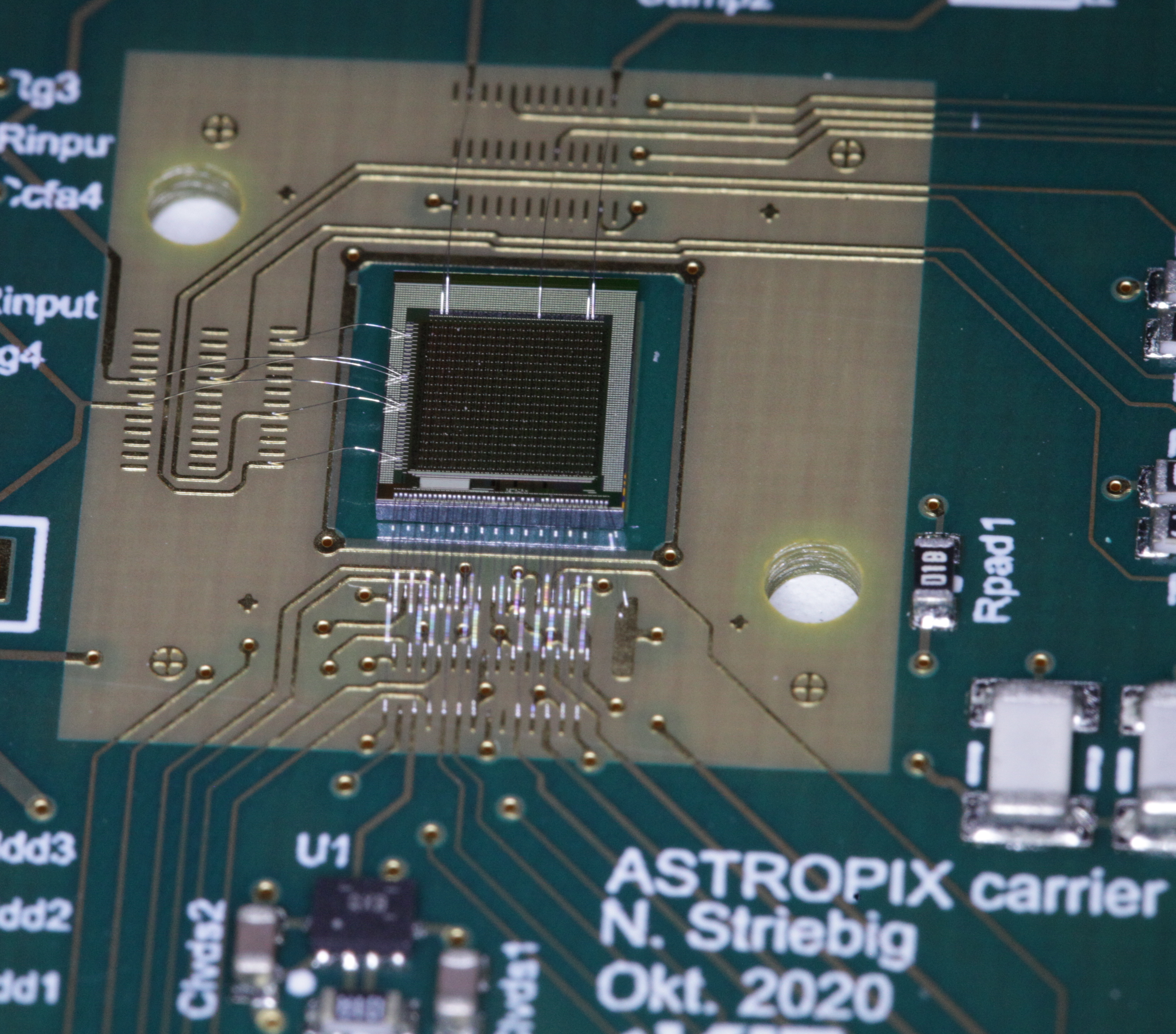}
        \caption{Picture of the AstroPix V1 detector on a carrier board.}
        \label{astropix_v1_picture}
    \end{subfigure}
\end{figure}

The next generation of AstroPix pixels, AstroPix V1 (pictured in \textcolor{black}{Figures \ref{astropix_v1_diagram} and \ref{astropix_v1_picture}}), has already been fabricated and is undergoing testing. The new V1 pixels are based off of the ATLASPix design, and include the amplifier and comparator all within the sensitive area. AstroPix V1 pixels are larger, with each pixel 200 $\mu$m by 200 $\mu$m (175 $\mu$m by 175 $\mu$m active area) and square. This adjustment to larger pixel area was made based on the simulations in Section \ref{sec:Sims}. AstroPix V1 is comprised of an 18 by 18 pixel matrix and includes 36 comparator outputs at the end of each row and column where the digitization occurs, as seen in Figure \ref{astropix_v1_diagram}. The digital output contains a 12 bit timestamp and a 10 bit value for the TOT. The analog output can be read from 36 pixels. This first generation of AstroPix pixels should give us a better understanding of how to further optimize the technology.

A design driver for the AstroPix technology is the power consumption, which is dominated by the amplifier within the pixel electronics. Initial tests of V1 have shown promising results approaching the required power usage listed in Table~\ref{goals_table}, with measurements of $\sim$20 mW/cm$^2$ with a 200 $\mu$m$^2$ pixel size \cite{nicolas}. With a larger pixel size of 500 $\mu$m$^2$, the pixel size for AstroPix V2, it is predicted we can reach under 1 mW/cm$^2$.

With AstroPix now the primary technology candidate for AMEGO-X, it is critical to rapidly optimize and prove the readiness of this technology. To this end, AstroPix V2 is already being designed and will include a 500 $\mu$m pixel pitch. Detailed simulations of the radiation environment in space were performed to determine the expected data rate of 1 MHz hits per pixel. This gives a more lenient resolution of 8 bits. AstroPix V2 also includes a 16 bit ToT resolution to optimize the energy response. The goal is to have V2 fabricated and undergo testing by the end of 2021. 

\section{Conclusion}

With a new generation of multimessenger missions on the horizon, it is critical to have technology that supports our science objectives. A revamped pair/Compton Si tracker technology that delivers on the driving parameters of energy resolution and power consumption is necessary to properly investigate the promising soft-gamma and medium-energy sky. The first step of AstroPix, characterizing the existing ATLASPix detector, has been an auspicious start to developing this technology. The intrinsic energy resolution of the ATLASPix detector, represented by the analog output, already exceeds our baseline AstroPix requirement of $<$10\% at \textcolor{black}{60} keV. The digital output of the detector results in an energy resolution that is insufficient for gamma-ray science, but this is a problem that can be redesigned. Currently, the digital ATLASPix capability is devoted to timing resolution instead of energy resolution, and in future generations of pixels, this capability can be reassigned. As a first step, the ATLASPix results are extremely encouraging and suggest re-optimized pixels can improve on digital resolution.

Due to the ATLASPix energy resolution performance, as well as the other potential benefits of monolithic Si pixels such as reduced power requirements and improved angular resolution, AstroPix-type pixels are now a technology candidate for the proposed AMEGO-X Si tracker. This an exciting time for new high-energy missions and multimessenger astronomy, and it is important to supplement these science goals with cutting-edge detector technology.

\section*{Acknowledgements}
I.B, M.N., and H.F. acknowledge the Center for Research and Exploration in Space Science and Technology II (CRESST II); the material is based upon work supported by NASA under award number 80GSFC21M0002. This research is supported by grants from the National Aeronautics and Space Administration Astophysics Research and Analysis Program (APRA) via Solicitation NNH18ZDA001N-APRA.  The AstroPix Team thanks Mathieu Benoit for advice and expertise that was critical in getting ATLASPix operational. Software for this project included NumPy \cite{numpy}, Matplotlib \cite{matplotlib}, and SciPy \cite{scipy}.


 \bibliographystyle{elsarticle-num} 


\end{document}